# Beyond CNNs: Exploiting Further Inherent Symmetries in Medical Images for Segmentation

Shuchao Pang[1], Anan Du[2], Mehmet A. Orgun[1], Yan Wang[1], Quanzheng Sheng[1], Shoujin Wang[1], Xiaoshui Huang[3], Zhemei Yu[4]

[1]Macquarie University, NSW, Australia
[2]University of Technology Sydney, NSW, Australia
[3]The University of Sydney, NSW, Australia
[4]Shandong Women's University, Shandong, China
shuchao.pang@hdr.mq.edu.au

## Abstract

Automatic tumor segmentation is a crucial step in medical image analysis for computer-aided diagnosis. Although the existing methods based on convolutional neural networks (CNNs) have achieved the state-of-the-art performance, many challenges still remain in medical tumor segmentation. This is because regular CNNs can only exploit translation invariance, ignoring further inherent symmetries existing in medical images such as rotations and reflections. To mitigate this shortcoming, we propose a novel group equivariant segmentation framework by encoding those inherent symmetries for learning more precise representations. First, kernel-based equivariant operations are devised on every orientation, which can effectively address the gaps of learning symmetries in existing approaches. Then, to keep segmentation networks globally equivariant, we design distinctive group layers with layerwise symmetry constraints. By exploiting further symmetries, novel segmentation CNNs can dramatically reduce the sample complexity and the redundancy of filters (by roughly 2/3) over regular CNNs. More importantly, based on our novel framework, we show that a newly built GER-UNet outperforms its regular CNN-based counterpart and the state-of-the-art segmentation methods on real-world clinical data. Specifically, the group layers of our segmentation framework can be seamlessly integrated into any popular CNN-based segmentation architectures.

## 1 Introduction

Medical imaging has been playing a critical role in the whole clinical diagnosis process, especially in today's era when the technology for medical imaging is developing rapidly [Ting *et al.*, 2018]. To support the clinical diagnosis process effectively, it is imperative to develop automatic and accurate medical image segmentation technologies for computer-aided diagnosis, with lower false positive rate and false negative rate. In particular, tumor segmentation in CT volumes is one of the most challenging tasks in medical image segmentation. Specifically, the main challenge stems from the following aspects: (1) the low contrast between tumors and their surrounding tissues with similar appearances; (2) the unpredictability of tumors in location, shape, size, number from different patients; (3) the intensity dissimilarity within different parts in a tumor; and (4) limited medical data and manual errors in pixel-level annotations.

Currently, a proven and effective approach to address the above challenges is the utilization of Convolutional Neural Networks (CNNs) [Zhang *et al.*, 2019]. Based on CNNs, many new segmentation models [Mou *et al.*, 2019] have been proposed, e.g., the patch-based model, the image-based model, the non-local- or the attention- or the context aggregation-based models, and the 3D CNN model. Leaving aside their pros and cons, all of these models have partially improved the performance of medical tumor segmentation by designing novel network architectures. Most importantly, the common trait among these advanced segmentation models is that their performance and reliability heavily depend on regular CNN operations. Nevertheless, as we observe, basic convolutional operations can only exploit translational invariance into regular CNNs, while ignoring further inherent symmetries in medical images. Taking Figure 1 as an example, the upper Res-UNet model based on regular CNNs cannot generate consistent predictions for tumor regions when rotating the same test slice.

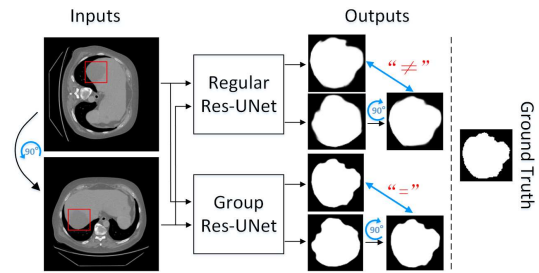

Figure 1: Equivariance visualization between regular CNNs and group CNNs on a test CT slice and its rotated version. Note that only segmented tumor regions are shown as outputs for clarity.

In order to equip regular CNNs with more symmetric properties, three representative solutions have been reported. First, the widely used data augmentation, as a common and effective method, has been proposed to obtain approximate invariance towards various transformations [Ciresan *et al.*, 2013]. Data augmentation can constrain the network training process as a whole to learn more filters for different transformations and thus improve performance, however, (Gap 1) *the size of learned CNN models grows bigger leading to higher redundancy of kernels and higher overfitting risk.* More importantly, (Gap 2) *this soft constraint does not guarantee invariance of well-trained CNNs on the test data or even on the training data.* Second, instead of implicitly learning symmetries by rotating input images, existing rotation equivariant networks with explicitly constraining feature maps can maintain multiple rotated feature maps at each layer and are easy to be implemented [Dieleman *et al.*, 2016]. Consequently, (Gap 3) *it largely increases memory requirements to save multifold feature maps by rotating and repeating original feature outputs at each layer.* Third, exploiting rotation equivariance by acting on filters has become a promising direction [Gens and Domingos, 2014; Marcos *et al.*, 2017]. Although rotating convolutional kernels can achieve local symmetries among different orientations at each convolutional layer, (Gap 4) *these solutions generally limit the depth and the global rotation equivariance of networks due to the dimensionality explosion and the exacerbating noise from orientation pooling operations.*

This paper proposes corresponding solutions to the above four gaps, inspired by group equivariant CNNs for general image classification [Cohen and Welling, 2016]. For Gap 1, *we build a symmetry group by incorporating translations, rotations and reflections together to significantly increase the utilization of each kernel and reduce the numbers of filters.* For Gap 2, *we design strong layerwise constraints to guarantee networks be equivariant at each layer with rigorous mathematical proofs.* For Gap 3, *we perform our equivariant transformations on filters, rather than feature maps, in order to reduce memory requirements.* For Gap 4, *our proposed group equivariant layers and modules can keep networks globally equivariant in an end-to-end fashion. Moreover, they can be stacked into deeper architectures for various vision segmentation tasks, with a negligible computational overhead over regular CNN-based counterparts.*

By integrating the above solutions, we propose a novel group equivariant segmentation framework by designing new additional group layers and modules with rigorous mathematical proofs, according to inherent symmetries in medical images. As shown in Figure 1, the bottom Group Res-UNet based on our proposed framework can consistently predict two liver tumor regions with rotated equivariant results and achieve more accurate segmentation performance. The main contributions of this work are summarized below:

- We propose novel group equivariant segmentation CNNs going beyond the concept of regular CNNs. Accordingly, more inherent symmetries existing in medical images are captured at each layer in more effective ways.
- The proposed group layers can significantly improve the degree of weight sharing and increase expressive capacity of segmentation networks. More importantly, the group layers of our novel framework can keep whole group CNNs globally equivariant for segmentation.
- This work reveals a common bottleneck of the current segmentation networks, that is, localizing tumors is easy, but delineating tumor boundaries is difficult. To mitigate this problem, all the proposed group layers can be easily generalized in these modern CNN-based segmentation architectures with better performance.

Our empirical study shows that (1) the proposed GER-UNet can achieve the state-of-the-art performance on the real-world clinical hepatic tumor data in different evaluation metrics; and (2) group segmentation CNNs can significantly outperform the baseline regular CNN-based counterpart.

## 2 Related Work

In this section, we briefly review equivariance of regular CNNs and introduce a new symmetry group concept. Then, data augmentation is discussed as a widely used approach. Furthermore, we discuss some works in rotation equivariant networks and novel group equivariant CNNs.

### 2.1 Equivariant Property & Symmetry Group

We usually regard CNNs having translation invariance because each weight-shared feature kernel can detect the same object that could appear in any position of the whole input image [LeCun *et al.*, 2015]. In other words, when shifting the input image, regular CNNs can give a corresponding shift in the output feature maps, which is also called translational equivariance. Nevertheless, many visual images, including medical ones, exhibit not only translation equivariance but also rotation and reflection symmetries as well [Veeling *et al.*, 2018]. The current CNN models lack such equivariant properties and have to learn more convolutional kernels with more training data to make up for the gaps. If we replace the traditional single translational symmetry by a symmetry group, which covers more equivariant properties such as translations, rotations and reflections, to convolute each input data, it would generate more powerful and predictable performance in various vision tasks. Therefore, the current regular CNNs can be regarded as a special case of group CNNs.

### 2.2 Data Augmentation & Rotation Equivariant Networks

Data augmentation is a widely used technique to train a more robust neural network model for real applications [Ciresan *et al.*, 2013]. In essence, data augmentation methods mainly rely on random transformations on original data to increase the amount of data. Although these simply augmented data do help improve performance, they require a larger model capacity to save such copies of each feature filter for all potential transformations. Actually, this soft constraint hardly guarantees equivariant properties on test data, even on training data, as shown in Figure 1. In addition, for the same problem, Dieleman et al. [2016] proposed a rotation equivariant network by directly rotating each feature map at every layer, but this would largely increase memory requirements. Recently, some rotated equivariant networks [Gens and Domingos, 2014; Marcos et al., 2017] rotate convolutional kernels to achieve local symmetries among all orientations. As a result, the network architectures are usually very shallow and the noise generated from intermediate layers makes it difficult to maintain global equivariance.

## 2.3 Group Equivariant CNNs

In a groundbreaking work [2016], Cohen and Welling introduced group equivariant convolutional neural networks by exploiting symmetry groups for general image classification. In particular, the group convolution operation can increase the representation capacity of the network without increasing the number of parameters. Based on this theory, there are a growing numbers of works for various computer vision applications [Veeling et al., 2018; Li et al., 2018; Winkens et al., 2018; Winkels and Cohen, 2019]. Among them, in terms of segmentation tasks, Veeling et al. [2018] directly utilized the classification network proposed in [Cohen and Welling, 2016] to distinguish each input patch cropped from the whole image for segmentation. Furthermore, Xiaomeng Li et al. [2018] proposed an automatic skin lesion segmentation method. Although the work proposes a deeply supervised learning model, it does not give an accurate interpretation with rigorous mathematical proofs. Similarly, a short paper [Winkens et al., 2018] proposes to improve semantic segmentation performance by exploiting rotation and reflection symmetries, but it lacks detailed method illustrations and sufficient experimental verification.

In contrast to above work, our proposed segmentation framework has extended the theory of group equivariance [Cohen and Welling, 2016] from image classification to semantic segmentation, by further devising group equivariant Up-sampling, Output and Skip Connection modules for feature decoding and fusion, with rigorous mathematical proofs. Moreover, our methodology belongs to image-based segmentation category by exploiting an efficient and end-to-end segmentation network acting on the whole image. In the next section, we demonstrate these group modules from our framework in detail, and then construct a novel Group Res-UNet for hepatic tumor segmentation. The implementation details will be made available online (*https://github.com/shuchao1212/GER-UNet*).

# 3 The Proposed Group Equivariant Segmentation Framework

This section first introduces mathematical convolution formulas for signal and image processing, and then discusses symmetric properties with kernel-based equivariant operations. Then, we propose several core modules by adding layerwise symmetry constraints into our framework. Finally, we design a Group Equivariant Res-UNet (named GER-UNet) as a simple example to illustrate how to make these core modules work together for medical tumor segmentation tasks with the global equivariance.

## 3.1 Mathematical Convolution

In mathematics, the convolution operation is a main tool for signal analysis and processing due to signal attenuation over time. For example, assume that an input signal function $f(t)$ and a time response function $g(t)$ are given, then the output signal at time $T$ is calculated as follows:

$$[f * g](T) = \int_{-\infty}^{+\infty} f(t)g(T-t)dt. \quad (1)$$

Therefore, we can observe that the convolution operation consists of two parts: a function rollover (from $g(t)$ to $g(-t)$) (including a further sliding ($g(T-t)$)) and an integral (or weighted sum). Based on this theory, current CNNs also exploit it from the continuous form to the discrete form:

$$[f * w_i^{(t)}](x) = \sum_{y \in \mathbb{Z}^2} \sum_{k=1}^{K^{(t-1)}} f_k(y) w_{i,k}^{(t)}(x-y), \quad (2)$$

where $f: \mathbb{Z}^2 \to \mathbb{R}^{K^{(t-1)}}$ is the input function at the $t^{th}$ layer which means that the stack of feature maps $f$ outputted at the $(t-1)^{th}$ layer returns a $K^{(t-1)}$ vector at each pixel coordinate $(u,v) \in \mathbb{Z}^2$; similarly, $w_i^{(t)}: \mathbb{Z}^2 \to \mathbb{R}^{K^{(t-1)}}$ is the $i^{th}$ convolutional kernel function at the $t^{th}$ layer. Therefore, for the translation equivariance of regular CNNs, we can see that the translation followed by a convolution is the same as a convolution followed by a translation [Cohen and Welling, 2016]:

$$\begin{aligned} \left[[L_t f] * w_i^{(t)}\right](x) &= \sum_{y \in \mathbb{Z}^2} \sum_{k=1}^{K^{(t-1)}} f_k(y-t) w_{i,k}^{(t)}(x-y) \\ &= \sum_{y \in \mathbb{Z}^2} \sum_{k=1}^{K^{(t-1)}} f_k(y) w_{i,k}^{(t)}((x-t)-y) \\ &= \left[L_t [f * w_i^{(t)}]\right](x), \quad (3) \end{aligned}$$

where $L_t$ is a translation operator by $y \to y + t$ and $f * w_i^{(t)}$ is also a function on $\mathbb{Z}^2$. However, this property from regular CNNs is not equivariant to a rotation operator $L_r$. On the contrary, this process has to be done by rotating the kernel $w_i^{(t)}$,

$$\left[[L_r f] * w_i^{(t)}\right](x) = \left[L_r \left[f * [(L_r)^{-1} w_i^{(t)}]\right]\right](x), \quad (4)$$

where it is shown that rotating the input feature maps $f$ and then convoluting with a filter kernel $w_i^{(t)}$ is the same as the rotation operator by $L_r$ of the convolution between the original input $f$ and the inverse-rotated filter kernel $(L_r)^{-1} w_i^{(t)}$.

Analogously, the process can also be done for reflection operators with the above formula. Therefore, to achieve these kinds of goals without additionally learning rotated and reflected copies of the same filter by utilizing more training data, we introduce group operations on symmetry groups to replace these conventional operations in regular CNNs, which can equip CNNs with more equivariant properties.

## 3.2 Core Modules in the Proposed Framework

**The Group Input Layer** In our framework, all network operations are based on the same symmetry group, which consists of translations, rotations by multiples of $\pi/2$ and reflections. So, all convolutional input comes into groups of $\|G = \{g\}\| = 8$, corresponding 4 pure rotations and their own roto-reflections. Among all group convolution operations $G$, only the first layer is applied on original input images, which is called the Group Input Layer. Therefore, in this $\mathbb{Z}^2 \to G$ convolution process, we convolute the input image with 8 rotated and reflected versions of the same kernel, e.g., $w_i^{(1)}$ in Equation (5). And the whole group input layer is visually demonstrated in Figure 2 where the input image is of size $448 \times 448 \times 3$ and the kernel size is $3 \times 3 \times 3$. In addition, stride=1 and padding=1 are also set to control the size of output like regular CNNs.

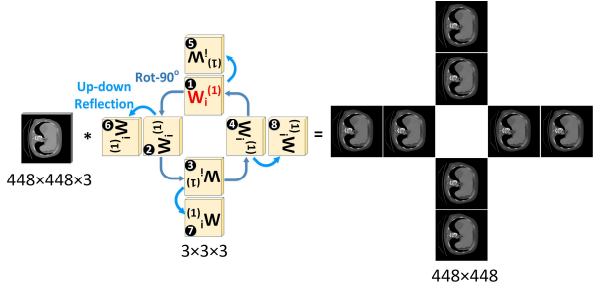

Figure 2: Visually convolutional representation of the Group Input Layer in our framework

$$[f * w_i^{(1)}](g) = \sum_{y \in \mathbb{Z}^2} \sum_{k=1}^{K^{(0)}} f_k(y) w_{i,k}^{(1)}(g^{-1}y), \quad (5)$$

where $K^{(0)}$ is the number of channels from input images. Note that, in this layer, the input image $f: \mathbb{Z}^2 \to \mathbb{R}^{K^{(0)}}$ and the filter $w_i^{(1)}: \mathbb{Z}^2 \to \mathbb{R}^{K^{(t-1)}}$ all belong to functions of $\mathbb{Z}^2$, but $f * w_i^{(1)}$ is a function on group $G$. And, the equivariance (under translations, rotations and reflections) of the group input layer can be derived by analogy to Equation (3), with any predefined group operators, e.g., $L_r: y \to ry$ as follows:

$$\begin{aligned}[] [[L_r f] * w_i^{(1)}](g) &= \sum_{y \in \mathbb{Z}^2} \sum_{k=1}^{K^{(0)}} f_k(r^{-1}y) w_{i,k}^{(1)}(g^{-1}y) \\ &= \sum_{y \in \mathbb{Z}^2} \sum_{k=1}^{K^{(0)}} f_k(y) w_{i,k}^{(1)}((r^{-1}g)^{-1}y) \\ &= [L_r[f * w_i^{(1)}]](g). \end{aligned} \quad (6)$$

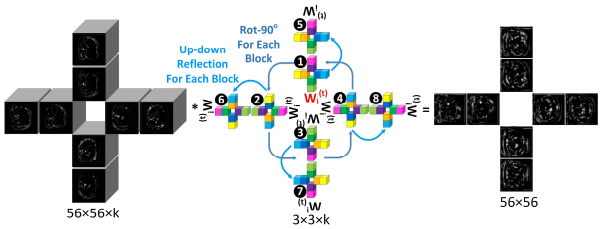

Figure 3: Visually convolutional representation of the Group Hidden Layer in our framework

**The Group Hidden Layer** Different with the group input layer, the next hidden layers are all operated on feature map groups based on the outputs from the previous layer. So, we call this $G \to G$ convolution process as the Group Hidden Layer. Because different kernels are designed for different orientations of input group feature maps, we need to convolute the input feature maps with 8 rotated and reflected symmetric group operations $G$ for each orientation. The details about this group hidden layer can be further understood in the following equation and the visually convolutional representation as shown in Figure 3. Therefore, all the layers and filters after the first group input layer are all functions on $G$.

$$[f * w_i^{(t)}](g) = \sum_{h \in G} \sum_{k=1}^{K^{(t-1)}} f_k(h) w_{i,k}^{(t)}(g^{-1}h). \quad (7)$$

Similar to Equation (5), the equivariance (under translations, rotations and reflections) of the group hidden layer can be derived with any predefined group operators $L_r: h \to rh$ as follows:

$$\begin{aligned}[] [[L_r f] * w_i^{(t)}](g) &= \sum_{h \in G} \sum_{k=1}^{K^{(t-1)}} f_k(r^{-1}h) w_{i,k}^{(t)}(g^{-1}h) \\ &= \sum_{h \in G} \sum_{k=1}^{K^{(t-1)}} f_k(h) w_{i,k}^{(t)}((r^{-1}g)^{-1}h) \\ &= [L_r[f * w_i^{(t)}]](g). \end{aligned} \quad (8)$$

**The Group Up-sample Layer** Like traditional upsampling operations in regular CNNs, e.g., FCNs [Long et al., 2015], we can also interpolate pixels into input feature maps with different modes such as the nearest and bilinear ones. In order to increase the size of outputs after each upsampling operation, we replace the traditional upsampling process only on an orientation by a Group Up-sample Layer over all 8 orientations. To this end, we could upsample symmetric group feature maps from each orientation respectively, or concatenate all group feature maps at the current layer and interpolate them all at once, and then separate them into 8 orientations again. Therefore, the upsampling operation is equivariant by acting on each group feature position from group equivariant feature maps. Note that this process is performed in sequence for keeping the group up-sample layer equivariant as well.

**Group Skip Connections** Skip connections have been proven effective in segmentation networks for recovering detailed features by combining encoder and decoder feature maps, for example, UNet [Ronneberger et al., 2015] and FCNs [Long et al., 2015]. Because of regular skip connections acting on two individual feature output blocks, we extend this operation on group outputs over all the orientations. In order to connect two sets of group convolutional outputs from the encoder and the decoder stages, we add or concatenate them together, following each orientation. In this way, our Group Skip Connections can obtain more details from each symmetry property, leading to more accurate segmentation predictions. Meanwhile, the sum of two group equivariant feature maps is also group equivariant.

**The Group Output Layer** This layer is the last layer in our framework to generate final segmentation score results, called the Group Output Layer. Furthermore, this is a $G \to \mathbb{Z}^2$ aggregation process, which aims to aggregate all the group segmentation outputs over various orientations. More importantly, this aggregation layer is a quite critical step to keep our group segmentation framework be equivariant over different rotations and reflections. To this end, we adopt globally average pooling over each orientation to obtain the equivariance for segmentation tasks. In our medical tumor segmentation tasks, we utilize the group output layer to transform all the orientation channels into a single pixel-wise 2D predicted output map.

$$f(x) = \frac{1}{\|G\|} \sum_{h \in G} f(h). \quad (9)$$

Note that we ignore the group max-pooling layer in our segmentation framework due to its significant reduction in resolution of feature maps, which is not conducive for segmentation [Fu *et al.*, 2019], using our group hidden layer with different strides instead. In addition, other batch normalization operations and non-linear pointwise activations (e.g., ReLU) are also locally equivariant on each symmetry group $G$, which can allow all these group layers to be stacked for much deeper group CNN segmentation models with global equivariance.

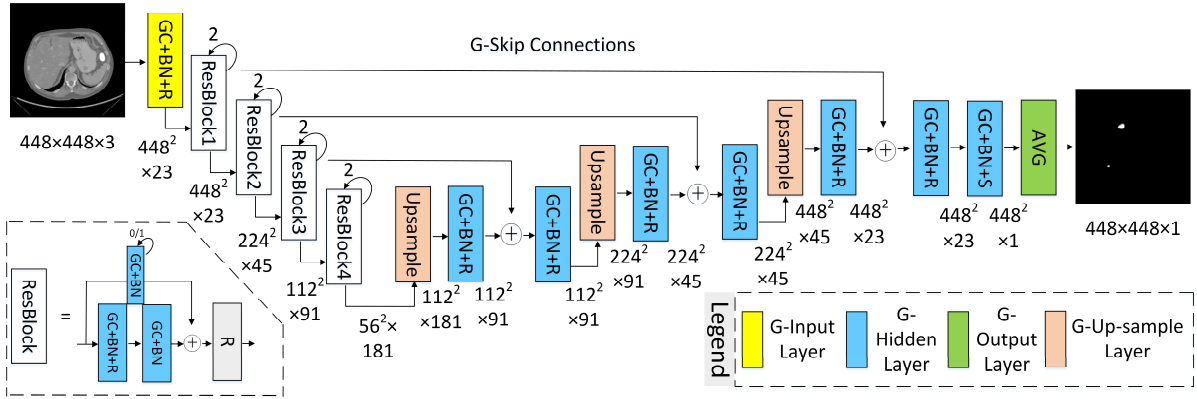

Figure 4: The architecture of our proposed Group Equivariant Res-UNet (named GER-UNet) for medical tumor segmentation

### 3.3 A GER-UNet Model

To validate our proposed group equivariant segmentation framework in improving segmentation performance, we design a GER-UNet model for medical tumor segmentation, which is based on ResNet blocks and our proposed core modules. The whole architecture of our proposed GER-UNet is shown in Figure 4, where all convolutions, batch normalizations, activation operations and other layers are constructed by our group equivariant counterparts. The architecture consists of 1 group input layer, 8 ResNet blocks, numerous group hidden layers, 3 group up-sample layers, 1 group output layer and others for the pixel-wise segmentation.

## 4 Experiments and Analysis

Extensive experiments are conducted on a real clinical hepatic tumor CT dataset, to answer the following questions: 1) is the whole segmentation framework equivariant to rotation? 2) can group equivariant CNN model outperform its regular CNN counterpart for segmentation? 3) is such a simple GER-UNet model better or more competitive than the state-of-the-art methods in this task?

### 4.1 Experimental Settings

**Datasets & Evaluation Metrics** To evaluate our proposed group equivariant segmentation framework, a challenging liver tumor segmentation dataset [Bilic *et al.*, 2019] is used in all experiments. This dataset includes 131 contrast-enhanced abdominal 3D CT scans, which are collected from 131 studied subjects who were suffering from different hepatic tumor diseases. Another challenge is that the images in this dataset come from different medical imaging acquisition devices around the world. In addition, among all CT slices, there are only 7190 slices with tumor information. To comprehensively compare tumor segmentation results among different segmentation methods, we use the most complete evaluation criteria in image segmentation tasks [Gu *et al.*, 2019; Schlemper *et al.*, 2019], including Dice, Hausdorff distance, Jaccard, Precision (called positive predictive value), Recall (called sensitivity coefficient or true positive rate), Specificity (called true negative rate) and F1 score. Among them, the smaller the Hausdorff distance, the better the segmentation results, for the others the opposite is the case.

**Parameter Setting** The dataset is randomly split into 4:1 for model training and testing. All experiments for all segmentation methods were conducted on Nvidia Tesla Volta V100. For our GER-UNet and the baseline regular R-UNet models, the batch size is set to 4 for training and the initial learning rate is 2e-4. The learning rate will gradually decrease as the number of training times increases. The training process is performed over 300 epochs, with an early stopping strategy for obtaining the optimal parameters. The widely used data augmentation techniques are also used to train the regular R-UNet model and others. In addition, the basic cross entropy loss function is used to compute the loss errors after each epoch. Also, we adopt the common Adam optimizer to update the whole network parameters. The other methods chosen for comparison are implemented following their papers and codes on the same dataset and settings with ours.

**Comparison Methods** We select 9 state-of-the-art segmentation methods, to compare with ours by considering the following representative perspectives: (1) U-Net and its variants: U-Net [Ronneberger *et al.*, 2015], Attention UNet [Schlemper *et al.*, 2019] and Nested UNet [Zhou *et al.*, 2018]; (2) Context Based Methods: R2U-Net [Alom *et al.*, 2019], CE-Net [Gu *et al.*, 2019] and Self-attention model [Wang *et al.*, 2018]; (3) Attention Based Methods: SENet [Hu *et al.*, 2018], DANet [Fu *et al.*, 2019] and CS-Net [Mou *et al.*, 2019].

### 4.2 Results and Discussion

All the experimental results are presented in Table 1. And all discussions and findings will be reported as follows according to the three questions raised before these experiments.

| | Methods/Metrics | Dice | Hausdorff Distance | Jaccard | Precision | Recall | Specificity | F1 |
|---|---|---|---|---|---|---|---|---|
| **The State-f-the-art** — *UNet-s* | U-Net [Ronneberger *et al.*, 2015] | 80.27 | 40.13 | 71.30 | 74.95 | 89.74 | 99.86 | 81.68 |
| | Attention UNet [Schlemper *et al.*, 2019] | 81.81 | 34.23 | 73.39 | 77.18 | 89.84 | 99.87 | 83.03 |
| | Nested UNet [Zhou *et al.*, 2018] | 80.14 | 36.65 | 71.29 | 74.79 | 89.43 | 99.86 | 81.46 |
| *Context* | R2U-Net [Alom *et al.*, 2019] | 78.75 | 35.43 | 69.27 | 72.84 | 89.56 | 99.83 | 80.34 |
| | CE-Net [Gu *et al.*, 2019] | 84.43 | 27.13 | 76.75 | 81.34 | 89.77 | 99.91 | 85.35 |
| | Self-attention [Wang *et al.*, 2018] | 83.52 | 31.56 | 75.83 | 81.16 | 88.37 | 99.91 | 84.61 |
| *Attention* | SENet [Hu *et al.*, 2018] | 83.42 | 30.45 | 75.58 | 79.95 | 89.44 | 99.90 | 84.43 |
| | DANet [Fu *et al.*, 2019] | 85.48 | 27.50 | 78.55 | 84.05 | 88.79 | 99.93 | 86.35 |
| | CS-Net [Mou *et al.*, 2019] | 84.12 | 26.42 | 76.21 | 80.23 | 90.63 | 99.90 | 85.11 |
| **Ablation Study** | Regular R-UNet | 82.60 | 32.44 | 74.77 | 79.82 | 88.52 | 99.90 | 83.95 |
| | GER-UNet (ours-w.-*add*) | 86.63 | 24.79 | 80.31 | 87.23 | 87.79 | 99.95 | 87.51 |
| | GER-UNet (ours-w.-*concat*) | 86.16 | 26.83 | 79.77 | 86.27 | 88.18 | 99.94 | 87.21 |

Table 1: Performance of segmentation methods on hepatic tumor segmentation dataset, measured by widely used evaluation metrics. Red numbers represent the best results and blue means the second best. Note that Hausdorff Distance uses pixel units and others %.

**Finding 1: Robust Equivariance of Our Proposed Segmentation Framework**

To evaluate the stability of predictions and the equivariance under rotation of the same input, we present a visual analysis shown in Figure 1. Although data augmentation techniques are used on each training image and different at each epoch, the well-trained regular Res-UNet gives very different predictions between an original test image and its rotated version, especially, for boundary regions. By contrast, the same Res-UNet architecture with our proposed group layers instead can be equivariant to the rotation operation on the test image, allowing us to obtain the same predictions by rotating the corresponding output score map. More importantly, by encoding these translation, rotation and reflection equivariances on each symmetry group $G$, the learned group CNN model can accurately discriminate object regions with much clearer boundaries, which looks more like the corresponding ground truth in Figure 1. In short, our proposed segmentation framework can keep learned group CNN model be equivariant for input transformations and largely improve the segmentation performance over its standard CNN counterpart.

**Finding 2: Overwhelmingly Superior to Its Regular CNN Counterpart**

As mentioned before, regular CNNs can be regarded as a special case of group CNNs because the former has only a single translation equivariance, whereas ours has more equivariant properties. As an ablation study, to fairly evaluate differences between them, we design a novel and standard UNet architecture based on ResNet, called regular R-UNet (or Res-UNet) and then we replace all these basic operations by our group equivariant layers (see Section 3). Meanwhile, in order to keep the model parameters consistent between them, we reduce the filter size of each layer to $1/\sqrt{8}$ due to 8 symmetry operations in our group $G$ for each filter, with total 12.78M (ours-w.-*add*) vs 12.75M parameters (Regular R-UNet). As shown in Table 1, we observe that our group equivariant segmentation model (GER-UNet) performs consistently better than its regular CNN version (Regular R-UNet). In particular, in terms of the important Dice similarity coefficient and Precision indexes, ours is 4.03 percent and 7.41 percent higher than its corresponding Regular R-UNet. This indicates that our novel group equivariant operations can significantly improve the performance of medical tumor segmentation in comparison to regular CNN-based layers.

**Finding 3: More Precise than the State-of-the-art Methods**

The current novel and popular segmentation methods not only build their network architectures deeper and wider based on UNet or FCNs, but also embed more advanced techniques into networks, as illustrated in the compared methods. To evaluate these state-of-the-art approaches and ours, we have trained and tested them on the same clinical medical tumor dataset. As shown in Table 1, results indicate that our proposed group equivariant GER-UNet performs consistently better than all the compared methods under different evaluation metrics for hepatic tumor segmentation, with the best Dice with 86.63%, Jaccard index with 80.31%, Precision with 87.23%, F1 score with 87.51% and the shortest Hausdorff Distance with 24.79 pixels. Overall, the average gain of our GER-UNet over all the compared methods (including the baseline Regular R-UNet) can achieve obvious improvements, which respectively are 4.18% in Dice, 7.40% in Hausdorff Distance, 6.02% in Jaccard, 8.60% in Precision, 0.06% in Specificity and 3.88% in F1 score. This illustrates that our group equivariant segmentation model can accurately capture hepatic tumor positions and give more refined tumor boundary delineations. In other words, it can also significantly reduce the rates of false positive and false negative results for early medical tumor detection and segmentation. The superior performance of our GER-UNet stems from the increased parameter sharing by encoding more robust symmetric operations for each convolutional filter. Meanwhile, we also observe that our group equivariant model has a faster convergence rate with fewer iterations (about 80 epochs) compared to others (about 300 epochs) based on standard CNNs. For another ablation study, we updated skip connec-tions by concatenating both feature map blocks from encoding and decoding stages, and it (ours-w.-concat) also achieves the best performance shown in Table 1.

In short, a simple GER-UNet has already fully demonstrated the potential of our proposed group equivariant segmentation framework. Moreover, we believe that the proposed group network framework and group operations would yield better results when deployed in deeper networks or combined with modern popular techniques (e.g., Non-local models, attention schemes and context aggregations).

## 5 Conclusions

To learn more precise representations for medical tumor segmentation, we have proposed a novel segmentation CNNs beyond regular CNNs by leveraging more inherent symmetries in medical images. To this end, we have developed kernel-based equivariant operations on every orientation, which can also guarantee whole segmentation networks being globally equivariant by encoding layerwise symmetry constraints at each group layer. This work not only dramatically reduces the redundancy of network filters, but also reveals a common bottleneck of current segmentation networks. Empirical evaluations on real clinical data have also shown the superiority of our novel group CNNs for medical tumor segmentation. With the proposed group up-sample layer, the group output layer and group skip connections, our group segmentation framework can be embedded into any popular CNN-based segmentation architectures.